\documentclass[pra,aps,graphics,a4paper,10pt,twocolumn]{revtex4-1}

\usepackage[margin=25mm]{geometry}

\usepackage{graphicx}
\usepackage{mathrsfs}
\usepackage{xfrac}
\usepackage{enumitem}
\usepackage{epstopdf}
\usepackage{amsmath}
\usepackage{amssymb}
\usepackage{braket}
\usepackage{color}
\usepackage{fp}
\usepackage{placeins}
\usepackage{array}
\usepackage{inputenc}
\usepackage{natbib}
\usepackage{hyperref}
\usepackage{tabularx}
\usepackage{tabu}
\usepackage{colortbl}
\usepackage[dvipsnames]{xcolor}
\usepackage{soul}
\usepackage{algorithmic}
\usepackage{ulem}
\usepackage{multirow}
\usepackage{bbold}
\newcommand{\flo}[1]{#1}

\newcommand{\robnote}[1]{}

\newcommand{\be}{ \begin{equation}}
\newcommand{\ee}{ \end{equation}  }

\newcommand{\nm}[1]{M_j}

\newcommand{\gd}{\Gamma_{\downarrow}}
\newcommand{\proj}[1]{\mathcal{P}_{#1}}
\newcommand{\A}[1]{B_{#1}}
\newcommand{\subs}[1]{H_{#1}}
\newcommand{\vecf}[1]{f_{#1}}
\newcommand{\dvecf}[1]{\dot f_{#1}}
\newcommand{\approxlvl}[1]{level~\ensuremath{#1}}
\newcommand{\Ab}{A}
\newcommand{\Em}{E}

\newcommand{\Imperial}{
Physics Department, Blackett Laboratory, Imperial College London, \\Prince Consort Road, SW7 2AZ, United Kingdom}
\newcommand{\ImperialCQD}{Centre for Doctoral Training in Controlled Quantum Dynamics, Imperial College London, \\Prince Consort Road, SW7 2AZ, United Kingdom}

\begin{document}

\title{Collective excitation profiles and the dynamics of photonic condensates}

\author{Benjamin T. Walker}\affiliation{\Imperial}\affiliation{\ImperialCQD}
\author{Henry J. Hesten}\affiliation{\Imperial}\affiliation{\ImperialCQD}
\author{Robert A. Nyman}\affiliation{\Imperial}
\author{Florian Mintert}\affiliation{\Imperial}

\begin{abstract}
Photonic condensates are complex systems exhibiting 
phase transitions due to the interaction with their molecular environment. 
Given the macroscopic number of molecules that form a reservoir of excitations, numerical simulations are expensive, even for systems with few cavity modes.
We present a systematic construction of molecular excitation profiles with a clear hierarchical ordering,
such that only modes in the lowest order in the hierarchy directly affect the cavity photon dynamics.
In addition to a substantial gain in computational efficiency for simulations of photon dynamics, the explicit spatial shape of the mode profiles offers a clear physical insight into  the competition among the cavity modes for access to molecular excitations.
\end{abstract}  

\maketitle

Photonic condensates~\cite{klaers2010,marelic2015, greveling2018} show intricate quantum many-body dynamics caused by the interplay of drive and dissipation.
They can be understood as analogues of atomic Bose-Einstein condensates~\cite{Anderson1995,Davis1995}, but they also show features, such as particle number fluctuations in time~\cite{schmitt2014observation}, that cannot be easily seen in their atomic counterpart.

Photonic condensates have many similarities with plasmonic lattice systems~\cite{hakala2018bose},
and they can also be described by rate equations \cite{keeling2016} derived from a fully quantized description of bosons and dye molecules~\cite{kirton2013}.
This model predicts condensation and mode competition at varying pump powers~\cite{PhysRevLett.120.040601} similar to behaviour in multi-mode lasers \cite{Tureci643}.

Due to pump and dissipation, these systems are typically out of equilibrium.
This is not only reflected in steady-state properties, but many interesting effects are likely to be transient, and observable only in experiments probing the temporal evolution.
Iconic examples in atomic Bose gases are memory effects~\cite{hofferberth2007non} or the absence of thermalization~\cite{kinoshita2006quantum}.
Recent experimental progress in the creation and control of photonic condensates opens the prospect of new experiments on their dynamics.

Corresponding theory can be based on existing models \cite{klaers2012statistical,keeling2016,milan} that have proven to be accurate descriptions of experiments~\cite{marelic2016, walker2018driven}.
In practice, however, such simulations will be limited by the effort required to describe the macroscopically large ensemble of dye molecules that is required for the photonic gas to be pumped and to thermalize.

Such limitations are the standard bottleneck whenever a small system interacts with a large environment.
As soon as assumptions that justify simplifying approximations ({\it e.g.} Markov) are not valid, one needs a well defined numerical framework in order to reduce the effort required for the description of the large environment.

Many approaches have managed to identify those environmental degrees of freedom that directly affect the system dynamics.
The rest of the environment can then be treated in an effective description 
\cite{iles2014environmental}, or with an identification of those environmental modes that directly affect the modes identified so far.
Such a construction can be based on a physically motivated notion of collective environmental modes~\cite{cederbaum2005short} or also on more abstract notions in the framework of hierarchical equations of motion~\cite{doi:10.1143/JPSJ.58.101,doi:10.1143/JPSJ.74.3131}.
Whereas most prior approaches begin such a construction with a microscopic system, we describe an identification of the most relevant environmental modes, based on the non-linear equations of motion~\cite{kirton2015,keeling2016} describing excitation of cavity modes and molecular excitation.
\flo{Our discussion is motivated by and focussed on photonic condensates, but readily generalises to different types of systems, in which a small number of system degrees of freedom interact with a large number of environmental degrees of freedom. Similar models are known to describe, for example, traffic jams, social networks or ecosystems~\cite{knebel2015}.}

\section{Model}

Typical experimental setups for photon condensation~\cite{klaers2010,marelic2015, greveling2018} include a multi-mode optical microcavity with curved mirrors which confine the light and induce an effective mass for the photons.
Since photons can easily be lost to transmission through the mirrors or scattering out of the cavity, the creation of photon condensates requires continuous pumping in order to compensate for these losses.
The cavity is thus filled with a large ensemble of dye-molecules that can absorb light from an external pump, and that enables the exchange of excitations between the different modes of the cavity via absorption and re-emission.
Due to the macroscopic number of dye molecules, the molecular excitation can be treated as continuous in space.
For the techniques developed in this paper, it is however more convenient to work in a discretised picture.
We will therefore define groups of molecules in small spatial regions, such that the coupling between all molecules in a group and any cavity mode is approximately constant within the group.
In principle, one can consider the limit of a continuum of groups, but for the sake of the efficient numerical framework that we are aiming for, it is more desirable to work with finite sums rather than integrals over continuous variables.

The starting point for our analysis is the set of rate equations \cite{keeling2016} for the set $\{n_i\}$ of occupations of the cavity modes and the excitation fractions for the molecules or groups of molecules.
Those equations can be put in a compact form if the molecular excitation fractions are understood as elements of a vector $f$.
The equation of motion for $n_i$ then reads
\begin{equation}
\begin{split}
\dot{n}_i = &(n_i (\Em_i+\Ab_i)+\Em_i) [Gf]_i-\gamma_i n_i\ ,
\label{eq:ndot}
\end{split}
\end{equation}
where $\Ab_i$ and $\Em_i$ are the rate of absorption and emission for the cavity mode $i$;
$G$ is a matrix with the elements $G_{ij}=g_{ij}\nm{j}$ comprised of the number $\nm{j}$ of molecules in the group
around position $j$ and the coupling constant $g_{ij}$ between cavity mode $i$ and a single molecule in this group.
The element $i$ of the vector $Gf$ is denoted by $[Gf]_i$, and
the overall decay constant $\gamma_i=\Ab_i \sum_j G_{ij}+\kappa$ also includes the cavity decay rate $\kappa$.

The rate equation for the molecular excitation reads
\begin{equation}
\begin{split}
\dot{f}&= \sum_{i=1}^{N_c} n_i\A{i}f-\A{0}f  +  v \ ,
\label{eq:fdot}
\end{split}
\end{equation}
in terms of the diagonal matrices $\A{i}$ (defined for $1\le i\le N_c$) with elements
\begin{equation}
\begin{split}
[\A{i}]_{pq}=&-(\Em_i+\Ab_i)g_{ip}\delta_{pq}\ ,
\end{split}
\end{equation}
the diagonal matrix
\begin{equation}
\A{0}=\sum_{i=1}^{N_c}\frac{\Em_i}{\Em_i+\Ab_i}\A{i}+(\gd+P)\mathbb{1}\ ,
\end{equation}
the vector
$v$ with elements
\begin{equation}
[v]_j=P+\sum_{i=1}^{N_c} g_{ij} \Ab_i \, n_i\ ,
\end{equation}
the pump rate $P$, and the decay constant $\gd$ for non-radiative decay or emission into free space.

Eqs.~\eqref{eq:ndot} and \eqref{eq:fdot} together provide a complete description of the system dynamics,
but typically only the dynamics of the cavity populations $n_i$ is of interest.
The number $N_c$ of cavity modes with non-negligible population is typically 
orders of magnitude smaller than the number of dye-molecules.
Even coarse-grained pictures in which many dye-molecules are grouped together require us to take into account a number $N_m$ of such groups that is orders of magnitude larger than $N_c$. 

\begin{figure}
	\centering
	\includegraphics[width=0.45\textwidth]{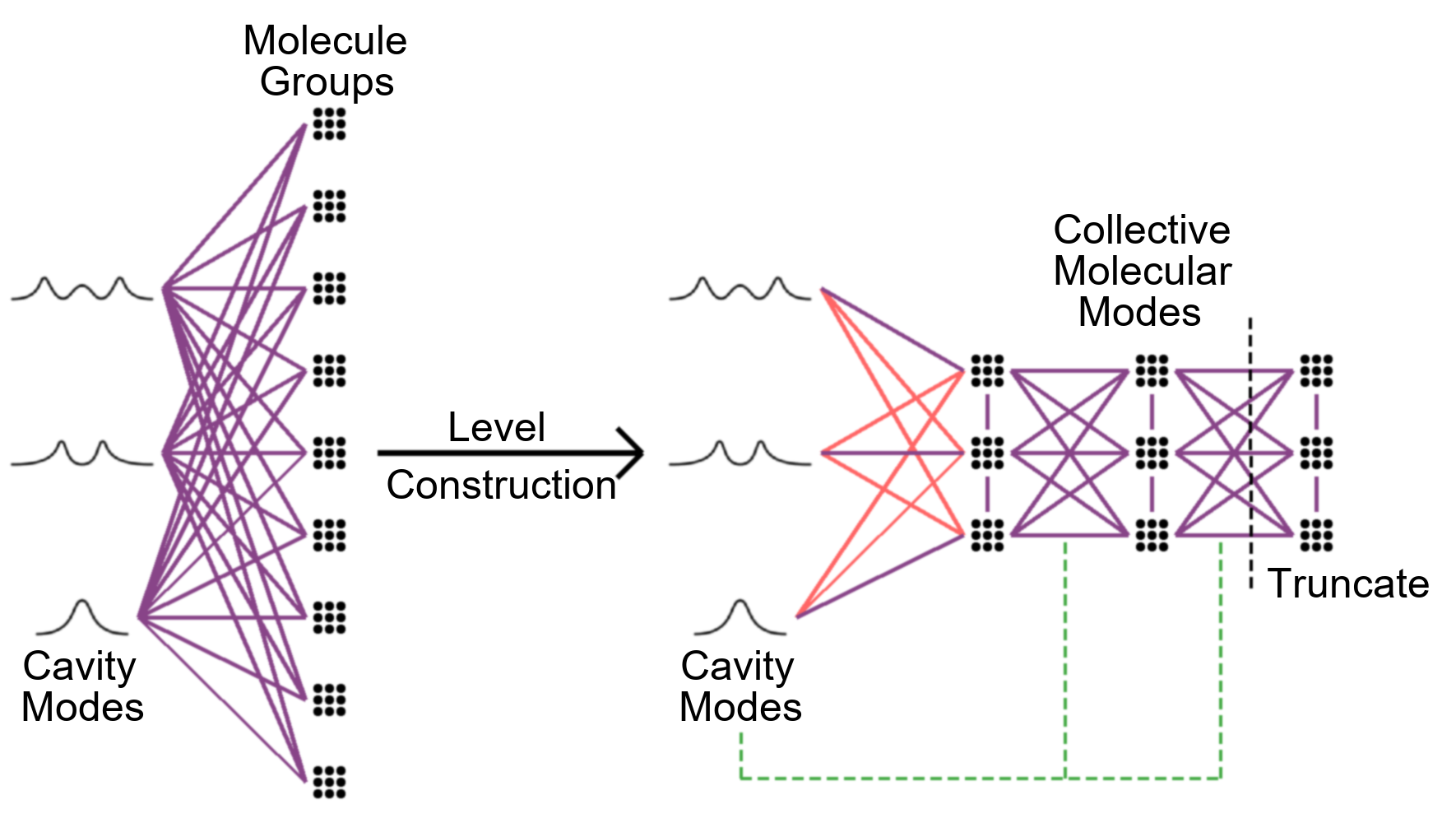}
	\caption{Schematic representation of the interaction between cavity modes and their molecular environment.
	With the environment described in term of the position of individual molecules or groups thereof (as depicted on the left), every cavity mode interacts with every molecule as indicated by purple lines.
	This requires taking into account the full molecular environment in a numerical simulation.\\
	With the construction of suitable excitation profiles (as depicted on the right), the interaction between the cavity modes and the environment is limited to a few selected modes.
	Those modes are considered to be of `level $0$' and modes of higher level affect the cavity dynamics only indirectly via their influence on the modes of the next lowest level.\\
	Quite importantly, the interactions between cavity and collective modes are not necessarily mutual: even though the profiles of non-lowest level do not directly affect the cavity dynamics, the dynamics of all excitation profiles are affected by the mode populations $n_i$ as expressed in Eq.\eqref{eq:fdot}, indicated by green lines.
	Similarly, red lines indicate the influence of mode populations on level $0$ modes without reciprocity.}
		\label{fig:schem}
\end{figure}

Our goal is therefore to define collective degrees of freedom of molecular excitation, and to identify the excitation profiles that are most relevant to the cavity dynamics, so that the dimension of the problem can be reduced substantially without significant loss in numerical accuracy.
More specifically, we will pursue a hierarchical construction as schematically sketched in Fig.~\ref{fig:schem}, such that only the lowest level in the hierarchy directly affects the cavity dynamics, and all higher levels affect the cavity dynamics only by affecting the next lowest level.

\section{Hierarchical construction}

Since the molecular excitation is described by an $N_m$-dimensional vector,
the following construction can be formulated in terms of mutually orthogonal subspaces $\subs{j}$ of this $N_m$-dimensional vector space and projectors $\proj{j}$ onto those subspaces.

The dynamics of mode population, described by Eq.~\eqref{eq:ndot}, shall depend only on vectors in $\subs{0}$.
The component $\vecf{j}$ of $f$ in subspace $\subs{j}$ ({\it i.e.} $\vecf{j}=\proj{j}f$), 
evolving according to Eq.~\eqref{eq:fdot},
shall depend only on vectors in the subspaces ranging from $\subs{0}$ to $\subs{j+1}$.
Such a construction will have the desired property that excitation profiles of high hierarchical level $j$ affect the cavity dynamics indirectly, so that a truncation after a reasonably low level promises an accurate, but efficient numerical description.

The matrix $G$ in Eq.~\eqref{eq:ndot} is an $N_c\times N_m$ matrix, and thus has at least $N_m-N_c$ null-vectors.
Since only vectors orthogonal to this null-space affect the cavity dynamics, one can directly see that $\subs{0}$ is given by the orthogonal complement to this null-space.
Equivalently, $\subs{0}$ is spanned by the $N_c$ $N_m$-dimensional vectors that $G$ is comprised of.
With the additional constraint, $[G{\bf e}_{i}]_j=\delta_{ij}$ for a set of mode-vectors ${\bf e}_{i}$ in $\subs{0}$, one also assures that the dynamics of $n_i$ depends on the molecular excitation only via the single profile ${\bf e}_i$.

Having identified $\subs{0}$, or any subsequent subspace $\subs{j}$, it is essential to ensure that the dynamics within this subspace depends on as small a number of additional degrees of freedom as possible.
The equation of motion for $\vecf{j}$
reads
\be
\dvecf{j}= \sum_{i=1}^{N_c} n_i\proj{j}\A{i}f-b\vecf{j}  - \proj{j}v \ ,
\ee
which can be understood as a linear differential equation for $\vecf{j}$ with a matrix ${\cal B}_j=\sum_i n_i\proj{j}\A{i}$.
Analogously to the construction of $\subs{0}$, it seems logical to identify the null-space of ${\cal B}_j$, since vectors from this null-space have no impact on the dynamics within $\subs{j}$.
Since ${\cal B}_j$ however depends on the cavity occupations $n_i$, which in turn are dynamical variables, we can only exclude those parts of the full vector space that belong to the null-space of ${\cal B}_j$ for all times.
Without prior knowledge of the dynamics, this can only be assured if one identifies the joint null-space of all matrices $\proj{j}\A{i}$.
The subspace $\subs{j+1}$ can then be defined as the orthogonal complement to the union of this joint null-space and all the subspaces $\subs{k}$ with $k\le j$.

With this construction, one satisfies the relation
$\proj{j}\A{i}\proj{k}=0$ for all $i$ and $k>j+1$,
or, equivalently
\begin{equation}
\begin{split}
x_j\A{i}x_k=0\hspace{.4cm} \mbox{for }&i,\ k>j+1\ ,\\
&\mbox{all vectors $x_j$ in $\subs{j}$ and} \\
&\mbox{all vectors $x_k$ in $\subs{k}$}\ .
\label{eq:matel}
\end{split}
\end{equation}
If this is given, then Eq.~\eqref{eq:fdot} results in the desired coupling between the excitation profiles of different hierarchical levels.
By construction, the dimension of $\subs{0}$ is no larger than $N_c$, and also the dimensions of all higher-level subspaces $\subs{j}$ are upper-bounded by $N_c^{j+1}$,
but, as we will see later on, the simple structure of the matrices $\A{j}$ in Eq.~\eqref{eq:fdot} results in a much more favorable scaling in $N_c$.

The definition of the hierarchy given so far includes an explicit description for its construction.
We can, however, arrive at a more efficient construction, exploiting specific properties of the coupling between the different levels of the hierarchy.
Since all the matrices $\A{i}$ are Hermitian, we have
\begin{equation}
x_j\A{i}x_k=(x_k\A{i} x_j)^\ast\ .
\end{equation}
Since the right-hand side vanishes for $j>k+1$, also the left-hand side vanishes,
or, equivalently we can conclude that $x_j\A{i}x_k=0$ for $j<k-1$ and all $i$.

This observation results in a substantial simplification of the dynamics of the excitation profiles:
originally, it was required that the dynamics of profiles in $H_j$ depend only on profiles $H_k$ with $k\le j+1$, but, in fact, it depends only on profiles $H_k$ with $j-1\le k\le j+1$.
The resulting equations of motion for the cavity occupations $n_i$ and the vectors $f_j=P_jf$ thus read
\begin{equation}
\begin{split}
\dot{n}_i& =(n_i (E_i+A_i)+E_i) [Gf_0]_i-\gamma_i n_i\ ,\\
\dvecf{j}&=\sum_{k=j-1}^{j+1}\sum_i n_i\proj{j}\A{i}\proj{k}f_k-b_0f_j  - \proj{j}v\ ,
\label{eq:dyn}
\end{split}
\end{equation}
where it is understood that the sum in the equation $\dot{f}_0$ includes only the term $k=0$ and $k=1$.
The operators $\proj{j}\A{i}\proj{k}$ describe linear maps from $H_k$ to $H_j$, {\it i.e.} subspaces with dimension substantially smaller than $N_m$.
Beyond this restriction to dynamics in low-dimensional subspaces, Eq.~\eqref{eq:dyn} also features only a few interdependencies between the different degrees of freedom:
the dynamics of $n_i$ is dependent only on $n_i$ and $f_0$, and the dynamics of $f_0$ is governed by $f_0$, $f_1$ and the $n_i$.
The dynamics of $f_j$ (with $j>0$) depends on $f_{j-1}$, $f_j$, $f_{j+1}$ and the $n_i$ as sketched in Fig.~\ref{fig:schem}.

This reduced coupling between the different subspaces $H_j$ not only results in a substantial reduction of numerical effort, but it also implies
that one obtains a valid hierarchy by requiring the condition $x_j\A{i}x_k=0$ (for $i\le N_c$) for $j<k-1$ instead of $j>k+1$, since the former condition necessarily guarantees the latter.

One can thus construct the hierarchy with the following prescription:
for a complete set of vectors $x_{jk}$ in $\subs{j}$ one can construct the vectors
$y_{ijk}=\A{i}x_{jk}$
for all $i\le N_c$ and all $k$.
Then one subtracts all components that lie in the subspaces $\subs{q}$ with $q\le j$,
{\it i.e.} one obtains the vectors $\tilde y_{ijk}=y_{ijk}-\sum_{q=0}^{j}\proj{q}y_{ijk}$.
Finally, $\subs{j+1}$ is the space spanned by all the vectors $\tilde y_{ijk}$.
Since this construction only requires the multiplication of matrices and vectors, and orthogonalisation of vectors, whereas a construction directly following Eq.~\eqref{eq:matel} would also require the construction of many null-spaces, this construction will generally be more efficient.

\section{Excitation profiles}

\begin{figure}[tb]
	\includegraphics[width=0.45\textwidth]{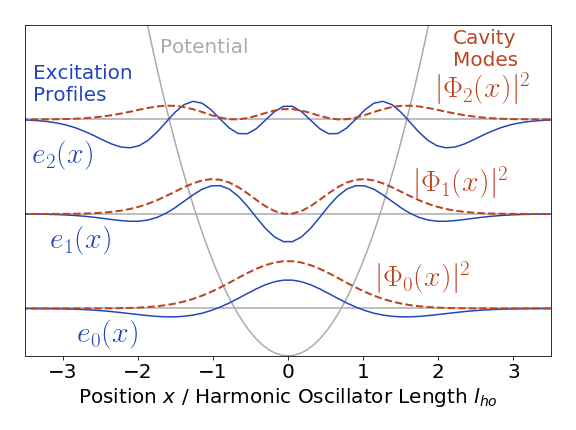}
	\caption{
	Moduli squared $|\Phi_i(x)|^2$ for $i=0,1,2$ of the three lowest eigenfunctions of a quantum harmonic oscillator are depicted in red.
	The corresponding harmonic potential and lines indicating the eigen-energies are depicted in grey.
	These functions describe well the profile of cavity eigenmodes.
	The excitation profile functions ${\bf e}_i(x)$ corresponding to each of these cavity modes are depicted blue.
	One can clearly see that each of the functions ${\bf e}_i(x)$ has minima, and even become negative, in spatial domains in which the cavity functions $|\Phi_j(x)|^2$ (with $j\neq i$) have maxima, such that the corresponding modes can couple strongly to the molecular environment.}
	\label{fig:molecularmodes}
\end{figure}

Besides the numerical benefits, the present construction also helps to understand the geometry of coupling between the cavity modes and the dye molecules.
Eq.~\eqref{eq:fdot} is based on the vector $f$ whose components describe the excitation fraction of the group of molecules around given positions.
Each excitation profile ${\bf e}_i$, on the other hand, corresponds to a spatial mode of excitation, and the shapes of these modes give valuable insight, as most easily exemplified with a one-dimensional system.

Fig.~\ref{fig:molecularmodes} depicts in red the moduli squared $|\Phi_i(x)|^2$ of the lowest three eigenfunctions of a quantum harmonic oscillator as representatives of mode functions of a cavity.
The coupling constants that determine the matrix $G$, and consequently the profiles ${\bf e}_i$ read $g_{ij}=|\Phi_i(x_j)|^2$, where $x_j$ is the position of the group $j$ of molecules.
The corresponding scaled profiles ${\bf e}_i(x_j)=[{\bf e}_i]_j$ are depicted by blue lines.

If there were only a single cavity mode, the function for the excitation profile would coincide exactly with the mode function of this cavity mode.
The differences between the profile and cavity functions reflect the competition among the cavity modes:
for example the profile ${\bf e}_0(x)$ has its maximum at $x=0$ due to the maximum of the mode function $|\Phi_{0}|^2$.
Further out in the tails, however, the profile ${\bf e}_0(x)$ becomes negative, reflecting the fact the the lowest eigen-mode of the cavity couples to the molecular environment in those areas more weakly than the higher frequency modes.

Similarly, the profile ${\bf e}_1(x)$ has a clear minimum around $x=0$, where symmetric modes -- in particular the ground state mode -- can strongly couple to the environment;
and it has additional, less pronounced minima in the tails, where this mode loses the competition against the next excited and further higher frequency modes.
The same features can also be observed in ${\bf e}_2(x)$, and higher frequency functions that are not shown here.

\section{Numerical accuracy}
\label{sec:num_acc}

The central question remaining to be answered is how accurate simulations with a truncation at a given \approxlvl{j} in the hierarchy are.
To this end, we consider a two-dimensional cavity such that the mode index $i$ is a double index $[m_x,m_y]$.
We take the mode-functions to be the eigenfunctions of a two-dimensional harmonic oscillator, consistent with a cavity with parabolically shaped mirrors. We take into account the lowest 4 energy levels with degeneracy $m_x + m_y + 1$, corresponding to 10 photonic cavity modes,
and we can express all system parameters in units of cavity decay constant $\kappa$ and the harmonic oscillator length $l_{ho}$.

We use 8 molecular groups per oscillation of the highest-frequency cavity-mode function in one spatial direction; fewer groups were found empirically to give inaccurate results.
A system of ten modes therefore requires at least 1521 molecular groups (i.e. $39\times 39$) which, along with a 2D molecular density of $10^{13}/l_{ho}^{2}$,
gives $\nm{j}=10^{12}$ molecules in each group.
For absorption and emission rate we use
$A_{[m_x,m_y]}=10^{-12}\kappa\ [1.83,\ 4.21,\ 10.3,\ 25.6]_{m_x+m_y}$, and $E_{[m_x,m_y]}= 10^{-10}\kappa\ [4.81,\ 5.69,\ 6.97,\ 8.31]_{m_x+m_y}$
consistent with Ref.~\cite{PhysRevLett.120.040601}.
The molecular decay rate is taken as $\gd =\kappa/4$.

\begin{figure}[tb]
	\centering
	\includegraphics[width=0.49\textwidth]{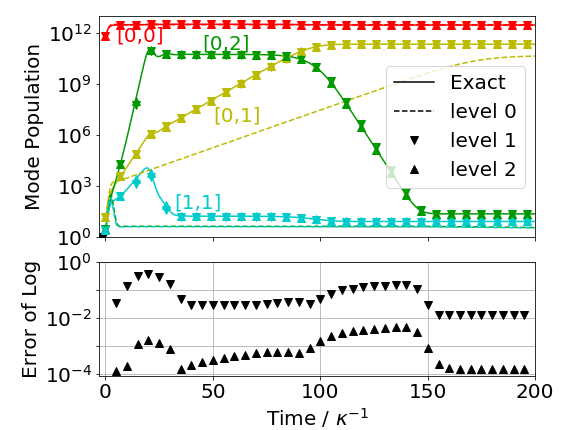}
	\caption{
The upper panel depicts the occupation of several cavity modes obtained with a numerically exact simulation (solid line), and simulations with truncation after \approxlvl{0}, \approxlvl{1} and \approxlvl{2} (dashed line, $\blacktriangledown$ and $\blacktriangle$.
Already truncation after \approxlvl{1} results in very good agreement with the exact dynamics.
The lower panel depicts the error $\epsilon_j$ defined in Eq.~\eqref{eq:error} on a logarithmic scale, 
and helps to verify that truncation after \approxlvl{2} results in an accuracy below the \%-range.}
	\label{fig:example}
\end{figure}

Fig.~\ref{fig:example} depicts an example for the dynamics of cavity occupations, obtained with a simulation without the hierarchical construction of excitation profiles, {\it i.e.} a numerically exact method, and with simulations using this construction and truncation after \approxlvl{j} with $j=0,1$ and $2$, which we shall refer to as the level-$j$ approximation.
The system is initialised in the steady state for a pump power of $6.58\times10^{-6}\kappa$, and the pump power is instantaneously quenched (increased) to $2\times10^{-5}\kappa$ in order to induce dynamics.

Fig.~\ref{fig:example} shows how this quench results in an initial growth of all the cavity modes, with the numerically exact data depicted by solid lines.
The lowest cavity mode $[0,0]$, depicted in red, is initially condensed (having a photon occupation much greater than 1), and quickly reaches its new steady state with a higher occupation.
Mode $[0,1]$ (depicted in yellow) condenses, but requires a time window of about $100/\kappa$ to reach its new stationary state; mode $[0,2]$ (green) condenses on a time-scale of $20/\kappa$, but decondenses as mode $[0,1]$ has reached its stationary occupation.
Mode $[1,1]$ (turquoise) shows cusp-like maximum, with subsequent decline to its stationary occupation.

Simulations in terms of excitation profiles and truncation after \approxlvl{1} or after \approxlvl{2} capture all these features very well; deviations from the exact dynamics are barely visible,
{\it i.e.}, a truncation after \approxlvl{1} already provides an excellent description.
In order to assess the accuracy of those approximations more quantitatively,
we can define the error
\begin{equation}
\epsilon_j = \max_i \left|\log_{10}\frac{n_i^{e}}{n_i^{(j)}} \right| \ ,
\label{eq:error}
\end{equation}
for the level-$j$ approximation,
where $n_i^{e}$ denotes the cavity occupations obtained with the numerically exact simulation and $n_i^{(j)}$ denotes the cavity occupations
obtained with truncation after \approxlvl{j},
and the cavity mode that yields the largest error is selected.

\flo{The lower panel of Fig.~\ref{fig:example} depicts this error for truncation after \approxlvl{1} and \approxlvl{2} as function of time, and provides a representative example of the accuracy of the present method.
As one can see,}
 the error for the \approxlvl{1} approximation is typically below $1/10$ (on the relevant scale between $0$ and $12$), but can exceed this threshold and reaches maximum a value of $0.44$;
the error for the \approxlvl{2} approximation is always below $1/100$ and typically substantially smaller.
The enhanced error for times in the interval between $20/\kappa$ and $35/\kappa$ can be attributed to the fast population growth in the mode $[1,1]$, and the slight enhancement of errors between $100/\kappa$ and $150/\kappa$ is due to the mode $[0,2]$, where a slight mis-estimation of the onset of the population drop can result in some inaccuracy.

Even a truncation after \approxlvl{0} -- the roughest possible approximation -- manages to capture the occupation of the mode $[0,0]$ surprisingly well, and it also reproduces the slow growth and saturation to a stationary value of the mode $[0,1]$ qualitatively correctly.
For the modes $[0,2]$ and $[1,1]$, however, it misses the growth of occupation, and fails to describe their dynamics, apart from a very short initial time-window.
Given the roughness of approximation, this is not necessarily a surprise, but it rather seems surprising that some features are indeed obtained correctly.

\flo{The quench used in the example of Fig.~\ref{fig:example} is sufficiently large to result in changes of mode populations over several orders of magnitude.
Fast and substantial changes are the features that are difficult to capture with a low-level approximation,
as one can see in the lower inset of Fig.~\ref{fig:example} that shows enhanced errors whenever there are substantial changes in the mode populations.
We found that these errors do not so much result from a failure in capturing the size of changes in the mode populations, but rather in a slight temporal offset to these transitions.
Since a small temporal offset on the time-axis results in a large error in mode population if they grow or decrease quickly, this explains the temporal dependance of the errors in the lower inset of Fig.~\ref{fig:example}.
Consistently with this, we found that errors are smaller for smaller quenches in pump power than used in Fig.~\ref{fig:example}, and they can become a bit larger for stronger quenches resulting in even more rapid growth of mode populations.}

\section{Numerical efficiency}

Having assessed the accuracy of the simulations with the different truncations,
we can now discuss the corresponding computational times.    
To this end we chose 33 pump powers between $10^{-3.5}\kappa$ and $10 \kappa$, geometrically spaced ({\it i.e.} with the logarithm of the pump power evenly spaced),
and we selected all unequal pairs of those pump powers as initial and final pump.
With the system initialised in the steady state corresponding to the initial pump power, dynamics is induced with an instantaneous quench to the final pump power.
We consider the dynamics until fractional deviations from the true steady state are below $10^{-6}$ for each mode.
For the resulting $33\times 32=1056$ situations, we can compare the CPU time required for the numerically exact simulation and the simulation with truncation after level~$j$ with $j=0,1,2$ and $3$ in the excitation profiles.

\begin{figure}[tb]
	\centering
	\includegraphics[width=0.49\textwidth]{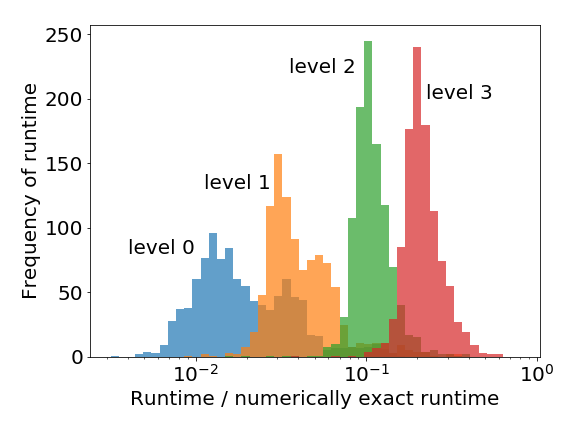}
	\caption{Histogram for the distribution of relative run times, {\it i.e.} ratio of computational time required for a simulation with truncation after \approxlvl{j} and computational time required for a numerically exact simulation, for 1056 simulations of cavity dynamics following a quench of pump. Data corresponding to $j=0,1,2$ and $3$ are depicted in blue, orange, green and red respectively.}
	\label{fig:runtime}
\end{figure}

The dimensions of subspace $\subs{0}$ for this system matches exactly the $10$ cavity modes, but, with
$\dim(\subs{1})=37$, $\dim(\subs{2})=79$ and $\dim(\subs{3})=110$, the subspaces for higher orders are substantially smaller than the general bound $\dim(\subs{j})\le N_c^{(j+1)}$ that is guaranteed by construction.
This favorable scaling in levels of the hierarchy is also reflected in Fig.~\ref{fig:runtime}, which depicts a histogram for the relative runtime, {\it i.e.} the ratio of computational time required for a simulation with truncation and the computational time required for a numerically exact simulation.

A simulation with a truncation after \approxlvl{0} requires taking into account 10 collective excitation profiles, as compared to $1521$ molecular groups in the exact simulation.
There is thus a reduction by two orders of magnitude in the number of dynamical variables,
and this reduction is neatly reflected by the typical relative runtime of about $10^{-2}$ as shown in Fig.~\ref{fig:runtime}.

The next level of approximation, {\it i.e.} truncation after \approxlvl{1}, requires keeping track of $37$ excitation profiles, and results in a typical speed-up of a factor of $30$;
even the worst cases in the tail of the distribution are still an order of magnitude faster than exact simulations.
Truncation after \approxlvl{2} with $79$ profiles still yields saving of computational time of about an order of magnitude, and a simulation with truncation after \approxlvl{3} with $110$ profiles requires only a factor of two more time than the next roughest approximation.

Independently of the accuracy sought in a simulation, one can thus see that the construction of suitable excitation profiles results in a substantial improvement of computational efficiency consistent with the number of dynamical variables taken into account.
Also, even though the rough level-$0$ approximation can at best make qualitatively correct predictions, its efficiency can make it a viable option for preliminary analyses.
If, for example, a scan over a large parameter-space is required in order to identify points of interest, such a rough but fast method can be used in order to decide which points are likely to warrant more thorough analysis.

\section{Pulsed Condensation}

\begin{figure}
	\centering
	\includegraphics[width=0.49\textwidth]{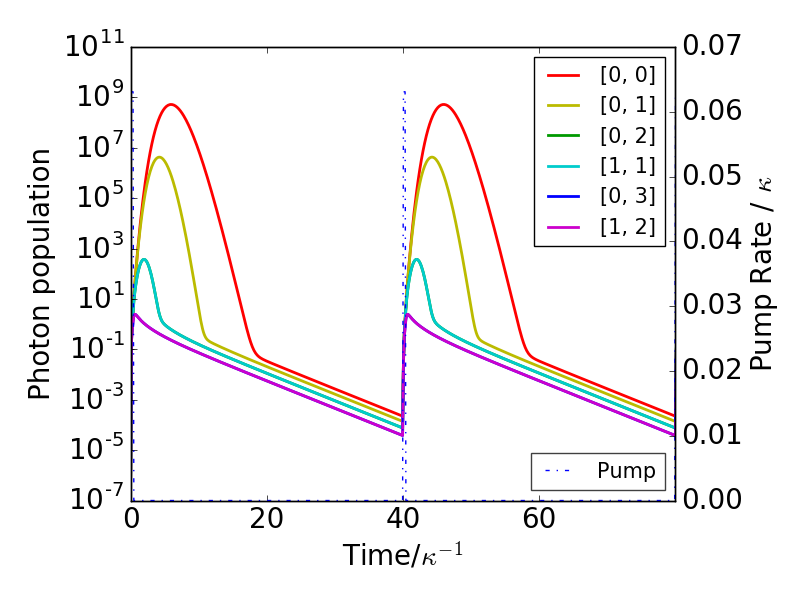}
	\caption{Time resolved population of cavity mode populations for pulsed pumping with a duty cycle of 0.01 and an average power of $10^{-3.2} \kappa$. In this case, modes which are degenerate but not symmetric to one another, such as [1, 1] and [0, 2] have the same population so lie directly on top of one another on the graph, unlike in figure~\ref{fig:example}.}
	\label{fig:pulsed_time_evolution}
\end{figure}

The rich physics that can be explored with this computationally efficient framework can be illustrated with the system behavior under pulsed pumping.
In fact, varying the duty cycle of a pulsed pump can have a dramatic effect on both the spectrum and the time averaged intensity of the intracavity light. Fig.~\ref{fig:pulsed_time_evolution} shows the evolution of the population in different cavity modes as a function of time over two time periods of pulsing for a duty cycle of 0.01 and an average power of 
$6.3\times 10^{-4}\kappa$.
The time period of $40\kappa^{-1}$, 
is long enough such that the system relaxes back to an equilibrium state between pulses,
but short enough for the system to show non-equilibrium behavior.
Besides this pulsed pumping,
Fig.~\ref{fig:pulsed_time_evolution} is based on the system and parameters described in section~\ref{sec:num_acc}.

As one can clearly see in Fig.~\ref{fig:pulsed_time_evolution}, all cavity mode populations begin to rapidly increase immediately after the pulse.
Since the pump is spatially uniform, all molecular excitation profiles obtain a similar occupation, and due to the strong pumping, they do so in a short period of time.
 Since the overlap of these excited molecules with different cavity modes depends on their position in space, all cavity modes have some molecules which emit into them more strongly than into any other cavity mode, leading to a large photon population in every mode. This is the process captured in the level 0 truncation of the equations of motion. Over longer timescales, spatial redistribution of excitations through emission and reabsorption starts to play a role in redistributing excitations between molecular reservoirs, as described in the level 1 truncation and higher. This process of redistribution between molecular reservoirs then allows the cavity ground state to develop a much larger population than the other excited cavity modes.

\begin{figure}
	\centering
	\includegraphics[width=0.49\textwidth]{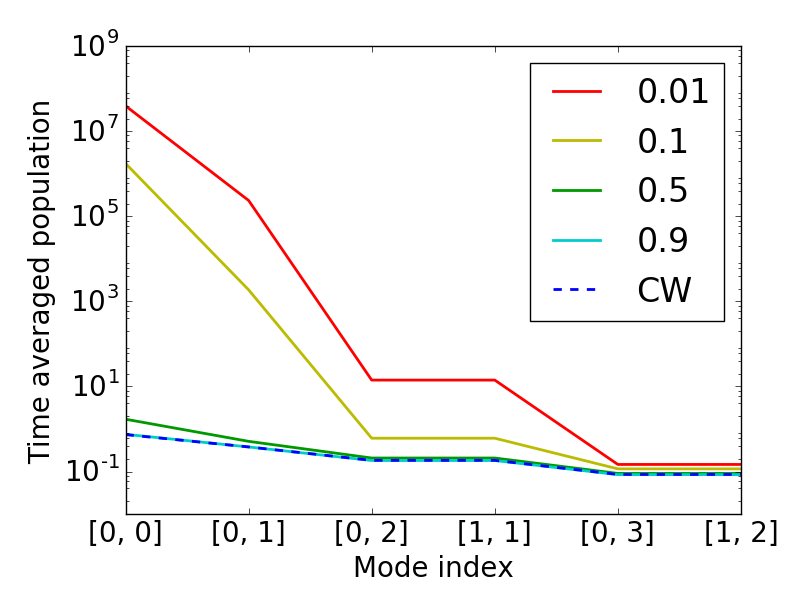}
	\caption{Time average population of cavity modes for different duty cycles, with average power fixed at $10^{-3.2} \kappa$ in a pulsed pumping scheme. Solid lines represent pulsed pumping with duty cycles shown in the legend. The dotted lines represents the results for continuous pumping at the same average power. Symmetric modes such as [0, 1] and [1, 0] were all considered in the simulation, but only one of the pair is plotted.}
	\label{fig:vary_dc}
\end{figure}

A further sign of this suppressed interaction
between excitation profiles can be seen in the relative populations of degenerate but spatially distinct modes, such as $[1, 1]$ and $[0, 2]$, in Figs.~\ref{fig:pulsed_time_evolution} and~\ref{fig:vary_dc}. While for long times and continuous pumping as shown in Fig.~\ref{fig:example}, two degenerate modes can develop different populations due to the difference in overlap of the molecular excitation profiles with other cavity modes, in the pulsed pumping regime these dynamics which appear at the higher order truncations of the equations of motion do not play a significant role. The populations of modes rather depends primarily on their absorption and emission coefficients as they interact with their molecular excitation profiles.

Figure~\ref{fig:vary_dc} shows how the spectrum and intensity of intracavity light varies with duty cycle of pulsing, from continuous pumping to a duty cycle of 0.01. By fixing the average power, the total number of excitations pumped into the system in one cycle is fixed. Pumping these excitations into the system in a short period of time favours stimulated emission, particularly in the lower energy modes which have greater emission rates relative to their absorption rates, leading to the dramatic change in spectrum. Since excited molecules can emit directly into free space at a rate of $\kappa / 4$, the total number of photons emitted from cavity modes into free space after a pulse of fixed energy can vary, depending on the balance of excitations between cavity photons and excited dye molecules, leading to the observed trend in average intracavity light intensity with duty cycle.

As well as providing insights into transient photonic condensates, these pulsed pumping schemes also give a method for experimental study of condensates previously inaccessible due to the requirements for high pump power. For example, experimental observation of decondensation under increasing pump power~\cite{PhysRevLett.120.040601} would require pump powers that cause photobleaching of dye molecules if such an experiment was performed with continuous pumping.
However with pulsed pumping, such an experiment could be performed with pumping that is orders of magnitude weaker, and thus substantially less challenging.
The behavior uncovered in Fig.~\ref{fig:vary_dc} thus suggests that many interesting transient phases and the study of their new physics will become experimentally accessible with a trick as simple as using pulsed instead of continuous driving.

\section{Conclusion}

Investigation into photonic condensates is moving towards transient, high power regimes, which require the full evolution of the cavity-mode populations to be properly understood.
With the construction of the excitation profiles and their hierarchical classification one can substantially reduce the numerical effort required for such analyses.
Simulations that are being enabled with this approach feature dynamical phase transitions \cite{slowPRL}, and pronounced changes in the equilibration time under changes in pump power.
Beyond the well-established phenomenon of slowing down of system dynamics close to such phase transitions, photonic condensates also feature this phenomenon far away from phase transitions where the system does not behave critically.

The construction of profiles of molecular excitation not only offers substantial numerical speed-up, but it also provides a very clear physical picture of mode competition that helps us to gain intuitive understanding of the non-linear dynamics of condensation and decondensation in photonic condensates.

The comparison in computational effort between approximate and exact simulations in this work was based on low-energy modes of cavities giving rise to harmonic oscillator eigenfunctions.
This allowed us to start with a rather coarse-grained description of the molecular environment with many molecules on a spatial group,
resulting in a reduced complexity of the underlying problem.
More strongly pumped systems and/or more complicated cavity geometries, as used in current and up-coming experiments \cite{flatten2016, 2017arXiv170706789D}, however do not admit such a straight-forward reduction of numerical effort.
Since the effort for the present approach scales only with the number of considered cavity modes, but is largely independent of the required spatial resolution, we believe that the construction of excitation profiles presented in this paper will enable many investigations of complex non-equilibrium many-body dynamics that otherwise would be prohibitively expensive in computational resources.

\section{Acknowledgement}

We are indebted to Andre Eckardt, Alex Leymann and Rupert Oulton for stimulating discussions.
Financial support through UK-EPSRC in terms of the Grants No. EP/ 312 J017027/1, No. EP/S000755/1 and the Centre for Doctoral Training Controlled Quantum Dynamics No. EP/L016524/1, and through the European Union's Horizon 2020 research and innovation programme under grant agreement No. 820392 (PhoQuS), is gratefully acknowledged.


\bibliographystyle{prsty}
\bibliography{citations}
\end{document}